\documentclass[final,3p,times,twocolumn]{elsarticle}

\usepackage{amssymb}

\usepackage{amsmath}
\usepackage{multirow}
\usepackage{dcolumn}
\newcolumntype{.}{D{.}{.}{-1}}

\usepackage{color}

\newcommand{\bma}[1]{\mbox{\boldmath${#1}\/$}}
\newcommand{\Cdot}{\bma{\cdot}}
\newcommand{\Nabla}{\bma{\nabla}}

\journal{Sensors and Actuators A: Physical}

\def\prb{Phys. Rev. B}

\begin{document}

\begin{frontmatter}



\title{Low-frequency noise characterization of a magnetic field monitoring system using an anisotropic magnetoresistance}


\author[Mateos,Lobo]{I. Mateos\corref{cor1}}
\ead{mateos@ice.csic.es}
\author[Mateos,Ramos]{J. Ramos-Castro}
\author[Mateos,Lobo]{A. Lobo\fnref{fn1}}

\cortext[cor1]{Corresponding author}
\fntext[fn1]{Deceased}

\address[Mateos]{Institut d'Estudis Espacials de Catalunya (IEEC), 08034 Barcelona, Spain}
\address[Lobo]{Institut de Ci\`encies de l'Espai (CSIC), 08193 Bellaterra, Spain}
\address[Ramos]{Departament d'Enginyeria Electr\`onica, Universitat Polit\`ecnica de Catalunya, 08034 Barcelona, Spain}

\begin{abstract}
A detailed study about magnetic sensing techniques based on anisotropic magnetoresistive sensors shows that the technology is suitable for low-frequency space applications like the eLISA mission. Low noise magnetic measurements at the sub-millihertz frequencies were taken by using different electronic noise reduction techniques in the signal conditioning circuit. We found that conventional modulation techniques reversing the sensor bridge excitation do not reduce the potential $1/f$ noise of the magnetoresistors, so alternative methods such as flipping and electro-magnetic feedback are necessary. In addition, a low-frequency noise analysis of the signal conditioning circuits has been performed in order to identify and minimize the different main contributions from the overall noise. The results for chip-scale magnetoresistances exhibit similar noise along the eLISA bandwidth ($0.1\,{\rm mHz}-1\,{\rm Hz}$) to the noise measured by means of the voluminous fluxgate magnetometers used in its precursor mission, known as LISA Pathfinder. 
\end{abstract}


\end{frontmatter}


\section{Introduction}\label{intro}

eLISA (evolved Laser Interferometer Space Antenna) is a space-based observatory proposed as a large space mission of the European Space Agency (ESA) and conceived to directly detect low-frequency gravitational radiation between 0.1\,mHz and 1\,Hz. This bandwidth, which is not observable from Earth, is expected to reveal some of the most exciting gravitational wave (GW) sources, such as massive black hole coalescence, compact binaries, and extreme mass ratio inspirals\,\cite{whitepaper}. eLISA will consist of three drag-free spacecraft in an equilateral triangle configuration with one-million-kilometer sides. The spacecraft constellation forms a two-link interferometer between freely floating  test masses (TMs) that act as the geodesic reference mirrors for the gravitational wave measurement. Hence, the changes between the TM distances caused by a GW shift the phase readout of inter-satellite laser interferometer. However, the weakness of GWs distort in a very small manner the geometry of space-time, therefore, measurements of exceedingly small displacements between the two TMs are necessary for their detection. 

For this reason, the environment surrounding the TM must be shielded from non-gravitational forces, which can perturb the geodesic motion of the bodies preventing the GW detection. Consequently, environmental conditions such as thermal, magnetic and random charging fluctuations need to be under stringent control\,\cite{bib:LISAReq}. Among the residual disturbance sources, a significant fraction is due to the magnetic environment created by the interplanetary  magnetic field, electronic units, and  other components of the satellite such as  the micro-thrusters and the solar panel cells. The non-gravitational  force  induced by the magnetic field ${\bf B}$ and its gradient is caused by  the  magnetic properties of the TM, i.e., the magnetization ${\bf M}$ and susceptibility $\chi$. This spurious force on the TM volume $V$ is given by\,\cite{bib:jackson}

\begin{equation}\label{equ:MagForce}
 {\bf F} = \left\langle\left[\left({\bf M} + 
           \frac{\chi}{\mu_0}\,{\bf B}\right)\Cdot\Nabla\right]{\bf B} \right\rangle V, 
\end{equation} 

\noindent where $\mu_0 = 4\pi\cdot10^{-7}\,{\rm m\,kg\,s^{-2}\, A^{-2}}$, and $\langle   \rangle$ denotes TM volume average of the enclosed quantity. This leads to keeping the magnetic background below certain values in order to ensure proper science operation of the GW observatory. Due to the important role that magnetic effects play in eLISA, a set of magnetic sensors will be placed in key locations with the purpose of quantitatively identifying the magnetic contributions that couple to the TM motion. To that end, magnetometers are aimed at reconstructing an accurate map of the magnetic field and gradient in the region occupied by the TM. 

The point of the magnetic sensing in eLISA has obviously been addressed first in its technology demonstrator called LISA Pathfinder\,\cite{bib:LPF}. As a consequence, the selection criteria to identify the applicable magnetometer technology is performed in view of the previous experience with the design of the LISA Pathfinder magnetic subsystem\,\cite{bib:DMU}, in which the selected scheme was a set of four tri-axial fluxgate magnetometers. This technology was chosen on grounds of its long heritage in space applications and the low noise along the LISA Pathfinder measurement bandwidth ($1\,{\rm mHz}\,\leq \omega/2\pi \leq\,30\,{\rm mHz}$). Looking toward eLISA,  a number of further improvements need to be taken into account\,\,\cite{bib:DDS_LTP,bib:MarcCQG}, which have derived in the study of alternative technologies to fluxgate magnetometers. Consequently, the main sensor characteristics that need to be addressed are: (i) compactness, so as to allow more of them to be incorporated in the spacecraft, and moreover, improve the spatial resolution; (ii) sufficiently low magnetic and thermal back-action effects on the spacecraft environment to avoid disturbances, so that they can be placed closer to the TM \,\cite{MateosBack}; (iii) low noise performance down to 0.1 mHz. In particular, the main purpose of this work is the development of a system capable of monitoring the slow drifts of the environmental magnetic field in eLISA by using chip-scale magnetometers.

Regarding the noise performance for the lower end of the eLISA bandwidth, magnetic field fluctuations across the TM are expected to be dominated by a time-varying interplanetary magnetic field not lower than $100\,{\rm nT \,Hz}^{-1/2}$\,\cite{bib:Bip1,bib:Bip2}, while the spacecraft's magnetic sources are expected to be the main contributors to the magnetic field gradient fluctuations\,\cite{bib:TMproperties}. Therefore, to be on the safe side, although eLISA requirements at subsystem level and the distribution of the magnetic sources in the spacecraft are still not formally defined, the noise performance of the magnetic measurement system should be at least one order of magnitude less noisy than the expected interplanetary magnetic noise to be measured. This implies a sensitivity in the measurement system of 
 
 \begin{equation}
 S^{1/2}_{B,\rm system} \leq 10 \,{\rm nT \,Hz}^{-1/2},\,\,\omega/2\pi = 0.1\,{\rm mHz}.
 \label{eq:magneticreq}
 \end{equation}
  
The work presented here has been proposed as well as part of the magnetic field monitoring system within the STE-QUEST mission concept\,\cite{bib:ste-quest},  a high-precision experiment of the weak equivalence principle using space atom interferometry. 

The reason for using anisotropic magnetoresistive sensors (AMR) as an alternative to the LISA Pathfinder scheme with fluxgate magnetometers, is their mass, size and power restrictions for space applications\,\cite{bib:Michelena1,bib:Michelena2,bib:ICL1}. Besides, the AMR-type HMC1001\,\cite{bib:honeywell} presents the lowest noise level among different commercial magnetoresistive sensors\,\cite{,bib:1/fAMR2}. Nevertheless, an important disadvantage of the AMR technology is the intrinsic $1/f$ noise that limits its use for applications requiring long integration time\,\cite{bib:1/fAMR}. Extensive research was conducted on this topic at frequencies between $0.1\,{\rm Hz}$ and $10\,{\rm kHz}$. However, to our knowledge, the noise performance of the sensor and its electronics has not yet been explored in the lower end of the eLISA bandwidth (0.1\,mHz). A recent work has shown a noise level of $\simeq 100\,{\rm nT\,Hz^{-1/2}}$ at $1\,{\rm mHz}$\,\cite{bib:ICL2}, which clearly exceeds the value in Eq.\,\eqref{eq:magneticreq}. For these reasons, in this paper we study the low-frequency noise behavior of a prototype based on magnetoresistive sensors with dedicated noise reduction techniques, which are necessary to achieve the envisaged magnetic noise level for eLISA\footnote{For a more demanding scenario, a parallel study was performed using an atomic magnetometer\,\cite{bib:atomiceLISA}.}. The paper is organized as follows. In Section\,\ref{sec:techniques} a brief overview of the noise reduction techniques is explained. In Section\,\ref{sec:analysis} we analyze the noise and thermal contributions of the sensor and signal conditioning circuits to the overall noise. The experimental results are presented in Section\,\ref{sec:results}, and finally, the main conclusions are drawn in Section\,\ref{sec:conclusions}. 

\section{Noise reduction techniques: flipping and electro-magnetic feedback}\label{sec:techniques}

As detailed further on in the text, the intrinsic noise characteristics specified by the manufacturer of the magnetoresistors\,\cite{bib:honeywell} are non-compliant with the requirements in Eq.\eqref{eq:magneticreq}. For this reason, different electronic noise reduction techniques need to be assessed in order to minimize the sensor noise level in the eLISA frequency band. This section describes the methods to be studied.

\subsection{Flipping}
\begin{figure}[ht!]
\centering   
\includegraphics[width=1.0\columnwidth]{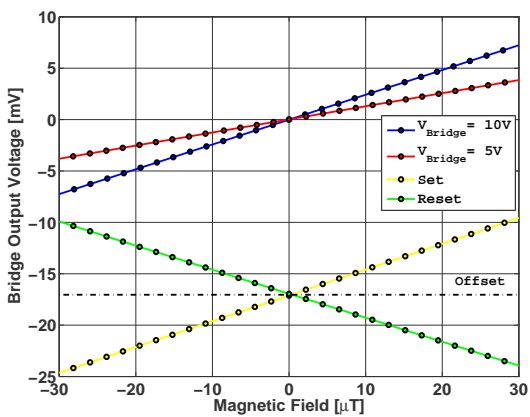}\\
\caption {Output characteristics as a function of the magnetic field after a {\em set} (yellow trace) and {\em reset} (green trace) pulse  with $V_{\rm bridge}=10\,{\rm V}$. Bridge offset extraction is performed for $V_{\rm bridge}=10\,{\rm V}$ (blue trace) and $V_{\rm bridge}=5\,{\rm V}$ (red trace).}
 \label{fig:sensitivity}
 \end{figure}

AMR sensors contain a thin film composed of a nickel-iron alloy with magnetic anisotropy. They have a sensitive axis to the magnetic field, the {\em hard} axis,  and another axis aligned with the sensor magnetization called the {\em easy} axis. Taking advantage of these properties, the flipping technique entails the periodic flip of the internal magnetization of the sensor strips by applying switching field pulses ({\em set}/{\em reset} pulses) generated by a thin film conductor, which is wound around the active area of the sensor\,\cite{bib:flipping}. The change of the magnetization direction induces the reversion of the output characteristic; as a result, the sensor output signal is modulated at the frequency of the switched pulses. Then, magnetic field measurements between each {\em set} and {\em reset} pulses are taken and subsequently demodulated. This sequence makes it possible to subtract the bridge offset, and its related temperature dependence, since the offset voltage remains unchanged while the sensor output reverses the polarity. Fig.\,\ref{fig:sensitivity} shows the opposite slopes in the output characteristics after the {\em set} and {\em reset} pulses, and the following offset voltage extraction for different bridge voltages. In addition, the main advantage of performing modulation techniques by using flipping pulses is the reduction of the $1/f$ noise within the desired bandwidth.  Another advantage is the recovery of the output signal degradation induced from strong external magnetic fields ($> 300\,\mu{\rm T}$), which resolves an important drawback of magnetometers that use ferromagnetic core, such as fluxgates.

\subsection{Electro-magnetic feedback}

In order to minimize the coupling between  temperature and magnetic field reading, the thermal dependence needs to be actively compensated during operation. Since the temperature drifts of the sensor sensitivity show up as a gain error in the measurement, a feedback controller is devoted to maintaining the bridge output close to zero, i.e. in balanced bridge condition, so as to reduce the thermal effects. By using electro-magnetic feedback, an integrated coil involved in the closed-loop controller induces an opposing field to counteract the field component detectable by the sensor. Then the current flowing through the compensation coil together with the current-to-field conversion of the coil give the strength of the magnetic field measurement. This method is particularly useful at low frequencies where temperature drifts become more significant in the overall sensor noise.

\section{Front-End Electronics}\label{sec:analysis}

\begin{figure}[ht!]
\centering
   \includegraphics[width=1.0\columnwidth]{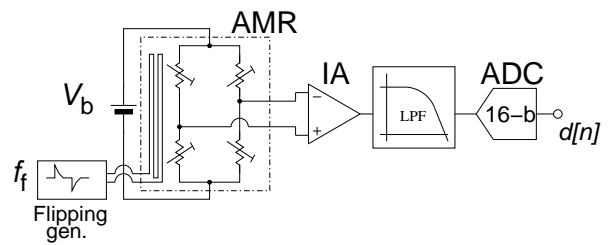}
   \caption {\label{fig:analog} Analog signal processing scheme for the flipping method.}
\end{figure}

The analog signal conditioning for the magnetic field sensing with the flipping method is shown in Fig.\,\ref{fig:analog}. The Wheatstone bridge is made up of four magnetoresistors, whose output signal is amplified, low-pass filtered, sampled and digitally demodulated.

The flipping generator circuit performs short {\em set/reset} pulses ($\tau\simeq 1\,\mu{\rm s}$) with peaks of 3.3\,${\rm A}$ along a strap of $1.5 \,\Omega$. Although the circuit delivers high current peaks, the duration of the pulses is so short that the energy stemmed from the charged-up capacitor is very small ($E=0.5\,C\,V^2 = 2.75\,\mu {\rm J}$ for  $C = 0.22\ {\rm \mu F}$). The flipping frequency $f_{\rm f}$ has been set to 
$5.5\,{\rm Hz}$, enough to reduce the $1/f$ noise of the instrumentation amplifier (IA) by modulating the signal from the magnetometer. The selected modulation frequency is a trade-off among the $1/f$ noise reduction, the effects on the magnetic and thermal disturbances produced by a more periodic switching signal, and the transient response after the pulses. Magnetic field measurements are acquired $10\,{\rm ms}$ after each {\em set} and {\em reset} pulses, so that all the flipping currents have died down below the micro-ampere level, and the low-pass filter settling time has elapsed. Therefore, glitches and transients in the immediate times after the flipping pulses are not  seen by the analog-to-digital conversion process.

\begin{figure}[ht!]
\centering
   \includegraphics[width=1.0\columnwidth]{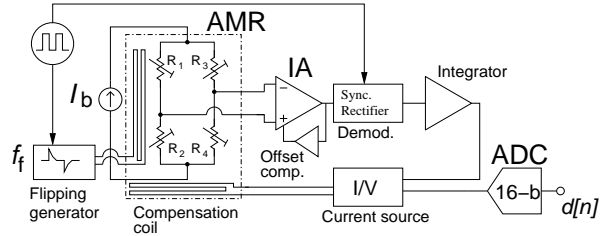}
   \caption {\label{fig:analog_feedback} Analog signal processing scheme for electro-magnetic feedback together with flipping.}
\end{figure}

The analog signal conditioning for electro-magnetic feedback together with the flipping method is shown in Fig.\ \ref{fig:analog_feedback}. The electro-magnetic feedback circuit is a closed-loop controller, in which a current regulator feeds the compensation coil with the measured magnetic field. In order to force the sensor output signal to zero, which is the remaining error, an integrator is required in the control loop. The measured field strength is represented as voltage, which is proportional to the compensation current.
 
\subsection{Low-frequency noise analysis}

The noise of the system can be split into two different parts, one coming from the intrinsic $1/f$ noise presented in the magnetometer itself and the other one coming from the signal conditioning circuit. In the following lines we describe the main noise sources in the whole system. 

The first stage of the circuit is the magnetic sensor constituted by a magnetoresistive Wheatstone bridge configuration. The nominal value of the resistors is $R_{\rm b,n} = 850\,\Omega$ and the bridge output sensitivity is $s_{\rm b} = s_{\rm AMR}V_{\rm b} = 136\ {\rm \mu V\ \mu T^{-1}}$, where $s_{\rm AMR} =  32\, {\rm \mu V\,V^{-1}\,\mu T^{-1}} $ is the AMR sensitivity and $V_{\rm b} = 4.25\ {\rm V}$ is the bridge voltage. Hence, the white-noise floor caused by the Johnson noise of the bridge resistances is $S^{1/2}_{{\rm AMR,}\ T} = (4k_{\rm B}TR_{\rm b,n})^{1/2} = 3.75\ {\rm nV\ Hz^{-1/2}}$ at $300\ {\rm K}$, where $k_{\rm B}$ is the Boltzmann constant. The voltage noise from the sensor is converted to equivalent magnetic field noise referred to the input dividing by the bridge sensitivity $S^{1/2}_{{\rm AMR,}\ T}/(s_{\rm AMR}V_{\rm b}) = 27.6\ {\rm pT\ Hz^{-1/2}}$. From this equation and assuming additive noise, the higher the sensor sensitivity or the bridge voltage the lower the equivalent magnetic field noise. At low frequencies, the corner frequency of the $1/f$ noise is around $60\ {\rm Hz}$, which leads to a sensor noise level of $21\ {\rm nT\ Hz^{-1/2}}$ at $0.1\, {\rm mHz}$. So, only the low-frequency contribution from the AMR is sufficient to exceed the requirements given in Eq.\,\eqref{eq:magneticreq}. As described further on in the text, apart from the intrinsic $1/f$ behavior, the contributions due to the thermal drifts \,\cite{bib:AMRtemp} deteriorate the noise performance. In order to overcome these limitations, the scheme shown in Fig.\,\ref{fig:analog_feedback} was implemented.

Concerning the noise contribution of the signal conditioning circuit, the apportionment of the bridge drive circuit is negligible compared with the sensor noise. Besides, {\em ratiometric measurements}\,\cite{Pallas} are performed in order to reduce drifts, noise or interference in the analog-to-digital conversion process. Hence, the voltage reference of the analog-to-digital converter (ADC) is also employed to drive the bridge. Afterwards, the bridge output is amplified by a space-qualified low-noise instrumentation amplifier (AD524) with a gain of $100\ {\rm V\ V^{-1}}$. The output noise introduced by this stage considering the closed-loop transfer function can be modeled as

\begin{align}
\nonumber
e^2_{\rm o,\ IA}&=\left[e_{\rm n,\ IA}^{2}\left(1+\frac{f_{{\rm c},\ e_{\rm n}}}{f}\right)+i_{\rm n,\ IA}^{2}\left(1+\frac{f_{{\rm c},\ i_{\rm n}}}{f}\right)R^{2}_{\rm b}\right]\\
&\times\left|\frac{K_{\rm eq} \cdot H_{\rm int}}{1 + H_{\rm oc} + K_{\rm eq}\cdot K_{\rm coil}  \cdot H_{\rm int} }\right|^2,
\end{align}
where $e_{\rm n,\ IA} = 7\ {\rm nV\ Hz^{-1/2}}$, $i_{\rm n,\ IA} =350 \ {\rm fA\ Hz^{-1/2}}$,  $f_{{\rm c},\ e_{\rm n}} = 3\ {\rm Hz}$  and $f_{{\rm c},\ i_{\rm n}} = 30\ {\rm Hz}$ are the input voltage/current spectral densities and their respective corner frequencies describing the noise characteristic of the IA, $K_{\rm eq}$ is the product of the bridge sensitivity $s_{\rm b}$ and the instrumentation amplifier gain, $K_{\rm coil}$ is the gain of the voltage-to-current converter ($4\,{\rm mA\,V^{-1}}$) multiplied by the compensation coil ratio ($1.96\,{\rm \mu T\,mA^{-1}}$), and $H_{\rm int}$ and $H_{\rm oc}$ are the integrator and offset compensation responses. The signal from the magnetometer, i.e., the input signal to the amplifier, is modulated by applying flipping pulses. As a result, the noise level of the IA is the one at the frequency of the modulating signal. For the AD524, the equivalent magnetic field noise with a modulating signal of only $5.5\ {\rm Hz}$ is  $64\ {\rm pT\ Hz^{-1/2}}$, thus fully compliant with the system requirements.

 \begin{figure}[ht!]
\centering
   \includegraphics[width=1.0\columnwidth]{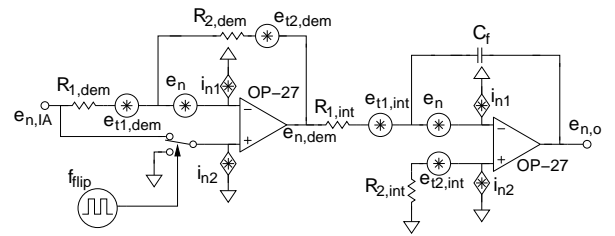}
   \caption {\label{fig:DemAndInt} Demodulator and integrator circuit with the main contributions considered for the noise estimation.}
\end{figure}

The phase sensitive detector is synchronized with the flipping pulses by using an analog switch, which alternates the sign of the unity gain amplifier in order to rectify the modulated signal. The noise contribution for both configurations (inverting and non-inverting amplifier) are not critical in the overall noise of the signal conditioning circuit. After the rectifier circuit the signal is demodulated and integrated (see Fig.\,\ref{fig:DemAndInt}). The output noise for the integrator is given by

\begin{align}
\nonumber
e^2_{\rm o, Int} &= \left[(e^2_{\rm n} + i^2_{\rm n}R_{\rm 2,int}^2 + e^2_{\rm t2,int})\left[1 + \left(\frac{f_{\rm i}}{f}\right)^2\right]\right. \\\nonumber
&\left.+ e^2_{\rm t1,int}\left(\frac{f_{\rm i}}{f}\right)^2 + i^2_{\rm n}R_{\rm 1,int}^2\left(\frac{f_{\rm i}}{f}\right)^2\right] \\
&\times\left|\frac{ 1 + H_{\rm oc}}{1 + H_{\rm oc} + K_{\rm eq}   \cdot K_{\rm coil} \cdot H_{\rm int}}\right|^2,\label{eq:intNoise} 
\end{align}

\noindent where $f_{\rm i}= 1/(2\pi R_{\rm 1,int} C_{\rm f})$, $e_{\rm t1,int}$ and $e_{\rm t2,int}$ is the thermal noise voltage of the resistor $R_{\rm 1,int}$ and $R_{\rm 2,int}$ (both of $10\,{\rm k\Omega}$), $e^2_{\rm n} = e^2_{\rm nf}(1+f_{\rm ce}/f)$ and $i^2_{\rm n} = i^2_{\rm nf}(1+f_{\rm ci}/f)$ are the amplifier input noise in terms of power voltage and power current density. The noise parameters of the op-amp are the noise floor ($e_{\rm nf} = 3\,{\rm nV\,Hz^{-1/2}}$ and $i_{\rm nf} = 0.4\,{\rm pA\,Hz^{-1/2}}$) and the corner frequency ($f_{\rm ce} =  2.7\,{\rm Hz}$ and $f_{\rm ci} = 140\,{\rm Hz}$). Hence, the equivalent output noise of the integrator is around $51\ {\rm pT\ Hz^{-1/2}}$ at $0.1\,{\rm mHz}$. 

In addition, an offset compensation integrator is also used between the output and the reference terminal of the AD524 in order to extract the offset of the modulated signal. The noise contribution of this circuit is not critical along the measurement bandwidth since the signal is still modulated at this stage.

\begin{figure}[ht!]
\centering
   \includegraphics[width=1.0\columnwidth]{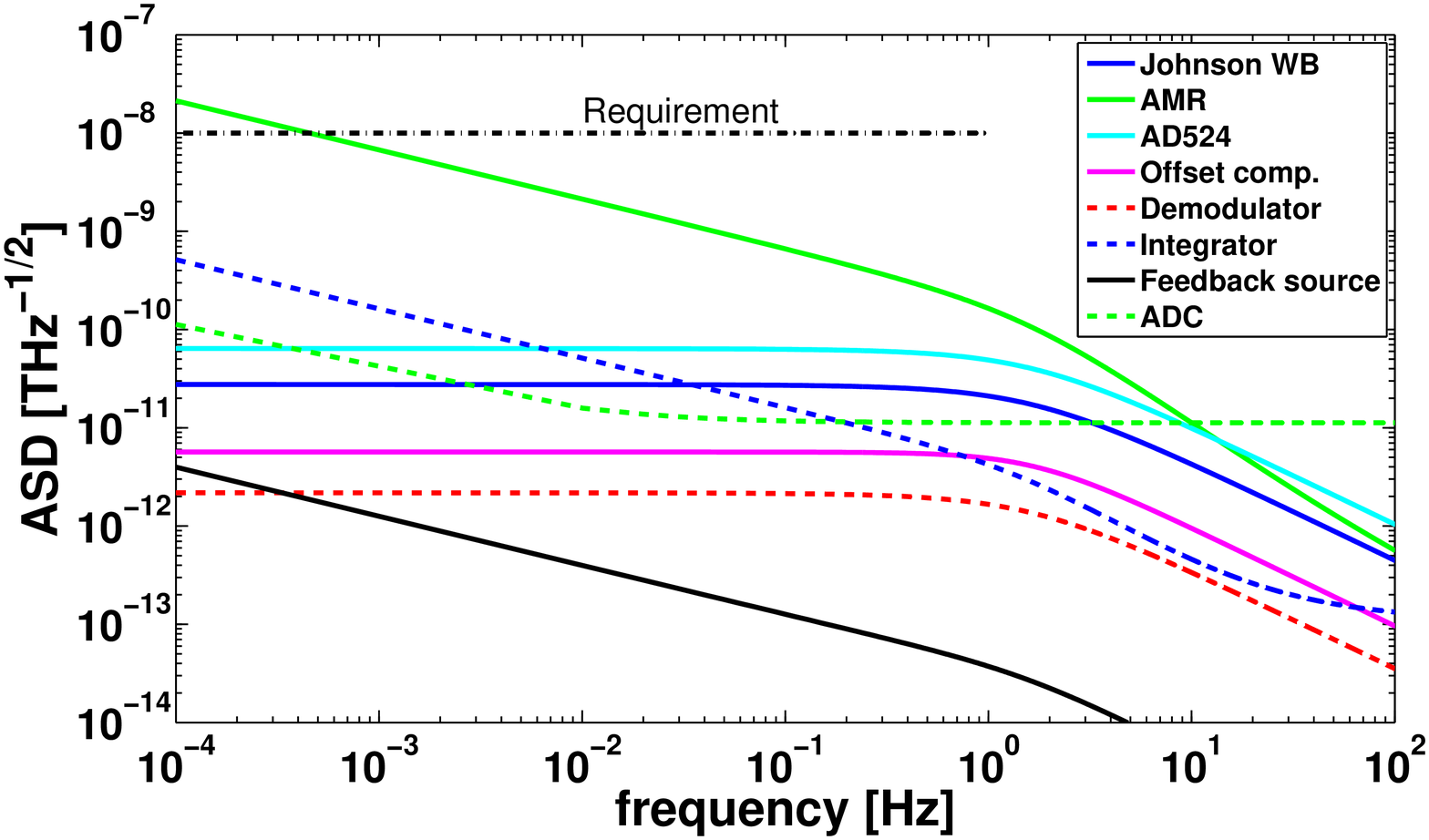}
   \caption {\label{fig:spectral_density_theor} Theoretical equivalent magnetic field noise of the signal conditioning circuit using flipping technique together with electro-magnetic feedback. Noise contribution of the AMR (green solid trace) shows the intrinsic noise of the sensor without flipping excitation. Its noise reduction due to the modulation technique is estimated experimentally in Section\,\ref{sec:results}. }
\end{figure}

The last stages in the closed-loop circuit are the  current source to drive the compensation coil and the ADC to measure the output voltage, which is proportional to the measured magnetic field. A {\em floating load} topology has been implemented on grounds of its simplicity and low-noise performance. Assuming a typical compensation coil ratio of $0.51\,{\rm mA\,\mu T^{-1}}$, the estimated equivalent magnetic field noise applied by the compensation source is $4\,{\rm pT\,Hz^{-1/2}}$ at $0.1\,{\rm mHz}$. This value is negligible compared with the intrinsic noise of the sensor itself ($21\ {\rm nT\ Hz^{-1/2}}$ at $0.1\, {\rm mHz}$, see Fig.\,\ref{fig:spectral_density_theor}). Regarding the ADC (ADS7809), the manufacturer gives a maximum rms noise of 1.3 LSB (least significant bit). This leads to a spectral noise density of $1.3 q/\sqrt{f_{\rm s}/2} = 1.4\,{\rm \mu V\,Hz^{-1/2}}$ ($11\,{\rm pT\,Hz^{-1/2}}$), where $q$ is the ADC voltage resolution for a 16-bit ADC with a full-scale range of $10\,{\rm V}$ ($\pm 5\,{\rm V}$). This contribution dominates over the ADC quantization noise $q/\sqrt{12fs/2} = 0.3\,{\rm \mu V\,Hz^{-1/2}}$ ($< 2.5\,{\rm pT\,Hz^{-1/2}}$). Since the low-frequency noise characteristics of the ADC are not given by the manufacturer, the corner frequency of the $1/f$ noise has been found at $10\,{\rm mHz}$ by an experimental fit to the data.

Fig.\,\ref{fig:spectral_density_theor} shows the theoretical output spectral noise density for the different stages of the signal conditioning circuit. As expected, the most important contribution at sub-millihertz frequencies is clearly the intrinsic $1/f$ noise of the AMR sensor, which is foreseen to be minimized in the experimental results by the flipping technique. We remark that although the $1/f$ noise of the sensor can be reduced, it can not be eliminated and is envisaged to continue being the dominant source in the overall noise of the system. In particular, the resistors in the bridge still suffer at long times from a $1/f$ behavior, and the AC modulation of the bridge output does not totally eliminate it.  On the other hand, the electronic noise sources from the signal conditioning circuits are  well below the magnetic requirement along the measurement bandwidth. This makes it possible to unveil the noise improvement of the sensor itself when utilizing the different noise reduction techniques. The IA, ADC, and Johnson noise of the magnetoresistances can limit the noise performance at frequencies higher than $1\,{\rm Hz}$, thus outside the eLISA bandwidth.

\subsection{Temperature coefficient analysis}\label{sec:TC}

The optimization of the thermal dependences in the sensors and the signal conditioning circuit is critical since slow temperature drifts may show up as a low-frequency noise contribution. One of the more sensitive elements to thermal changes is the resistors forming the Wheatstone bridge. Assuming the worst-case condition, the temperature coefficient (TC) of the bridge output voltage is

\begin{align}
 \alpha_{\rm b} &\simeq 2V_{\rm b}\left[\frac{R_1  R_2 }{(R_1+R_2)^{2}}+\frac{R_3 R_4}{(R_3 +R_4)^{2}}\right]\cdot \alpha_{\rm R, AMR}\\     
 &= V_{\rm b}\left(1 - \frac{\Delta R_{\rm b}^2}{R^2}\right)\cdot\alpha_{\rm R, AMR} \simeq V_{\rm b}\cdot\alpha_{\rm R, AMR},
 \end{align}

\noindent where $\alpha_{\rm R,AMR} = 0.25\,{\rm \%\,K^{-1}}$  is the TC of the AMR. The bridge resistances $R_{\rm b,n}$ change by an amount $\Delta R_{\rm b} = s_{\rm AMR}R_{\rm b,n}B = \pm 1.088\,\Omega$ for  a sensor range of $\pm 40\,{\rm \mu T}$. $R_1 = R_4 = R_{\rm b,n}- \Delta R_{\rm b}$ and $R_2 = R_3 = R_{\rm b,n} + \Delta R_{\rm b}$  are the magnetoresistive components shown in Fig.\,\ref{fig:analog_feedback}. Therefore, the maximum TC of the Wheatstone bridge is $10.6\,{\rm mV\,K}^{-1}$ and the equivalent magnetic field noise is given by

\begin{equation}\label{eq:tcbridge}
 S^{1/2}_{B,\rm \ WB}(\omega) = \frac{\alpha_{\rm b}}{s_{\rm AMR}\cdot V_{\rm b}}\cdot S^{1/2}_{T \rm , \ AMR}(\omega),
\end{equation}

\noindent where $S^{1/2}_{T \rm , \ AMR}$ is the thermal fluctuations in the magnetometer location. The thermal environment is not yet determined for eLISA, but the temperature fluctuations inside the satellite are expected to be lower than those required for LPF ($S^{1/2}_{T \rm , \ LPF} < 0.1\,{\rm K\,Hz}^{-1/2}$).  This leads to an equivalent magnetic field noise of $7.8\,{\rm \mu T\,Hz}^{-1/2}$, which is much larger than the noise level defined in Eq.\,\eqref{eq:magneticreq}. In view of the high thermal dependence of the AMR, flipping and electro-magnetic feedback are used to reduce the thermal drift effects in the sensors.

With the flipping scheme,  each magnetic readout is the average difference between two consecutive measurements with opposite polarization. Thus, the effect due to the temperature changes is now given by

\begin{equation}
 \alpha_{\rm b} \simeq V_{\rm b}\frac{\Delta R_{\rm b}}{R_{\rm b,n}}\alpha_{\rm R, AMR},
 \end{equation}

\noindent where $\alpha_{\rm b}$, assuming the worst-case condition and a full unbalanced Wheatstone bridge (full-scale range), is reduced to $13.6\,{\rm \mu V\,K}^{-1}$ ($0.1\,{\rm \mu T\,K}^{-1}$). Then, the equivalent magnetic field noise given by Eq.\,\eqref{eq:tcbridge} is $10\,{\rm n T\,Hz}^{-1/2}$, which barely achieves the requirements. 

\begin{figure}[ht!]
\centering
   \includegraphics[width=1.0\columnwidth]{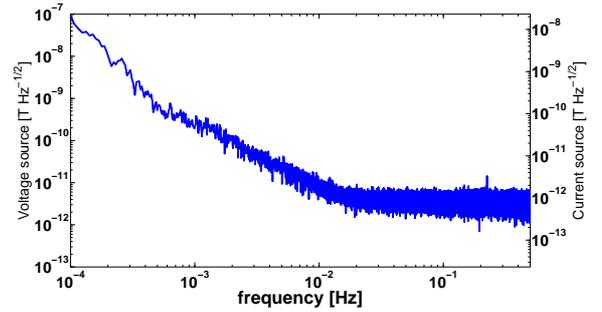}
   \caption {\label{fig:ASDTCAMR} Equivalent field noise contribution due to thermal fluctuations in the laboratory when using flipping technique at full-scale field range. Thermal contributions were estimated with constant voltage source ($S^{1/2}_B =  0.1\cdot10^{-6}\cdot S^{1/2}_{T \rm , \ AMR}(\omega)$, left vertical axis) and constant current source ($S^{1/2}_B =  24\cdot10^{-9}\cdot S^{1/2}_{T \rm , \ AMR}(\omega)$, right vertical axis) for the bridge excitation (see text for details).}
\end{figure}

When a constant voltage source feeds the bridge, the temperature dependency of the bridge resistance will vary the bridge output as $V_{\rm o} = V_{\rm b}\Delta R_{\rm b}/(R_{\rm b}(1 +  \alpha_{\rm R,AMR} T))$ for a worst-case error. By contrast, temperature sensitivity can be improved by using a constant current source instead, since variations in the resistances are partly compensated with changes in the voltage across the bridge. Then, the sensor output is equivalent to the constant current $V_{\rm o} = I_{\rm b}\Delta R_{\rm b}$ and the thermal stability is improved. Nevertheless, the AMR sensitivity also changes with temperature due to the energy-band structure of the magnetic material\,\cite{bib:AMRtemp1}. The error due to the temperature dependence of the sensor sensitivity is then $I_{\rm b}\alpha_{\rm sens}\Delta R_{\rm b} = 3.3\,{\rm \mu V\,K}^{-1}$ ($24\,{\rm n T\,K}^{-1}$), for $\alpha_{\rm sens} =  0.06\,{\rm \%\,K^{-1}}$ and a full-scale range of $\pm 40\,{\rm \mu T}$.  Thermal fluctuations during laboratory measurements are around $1\,{\rm K\,Hz^{-1/2}}$ at $0.1\,{\rm mHz}$. Thus,  as shown in Fig.\,\ref{fig:ASDTCAMR}, temperature dependences appear as additional noise at low frequencies when using exclusively the flipping technique.  As explained before, this effect is reduced by using a negative closed-loop that follows a null $\Delta R_{\rm b}$  to keep the bridge balanced. As a result, the gain errors barely affect the measurement.  

The TCs for the different stages of the circuit are compared in Table\,\ref{tab:TCs}, where thermal drifts  of the operational amplifier parameters (bias current, offset current, and offset voltage) can be neglected. On the whole, temperature dependences of the sensor, more precisely the TC of the magnetoresistance and bridge sensitivity, are the largest thermal contributors of the system.

\begin{center}
\begin{table}[ht!]
\caption{Temperature coefficients referred to the input for the stages of the electronics in which flipping and electro-magnetic feedback techniques were used. $I_{\rm b}\Delta R_{\rm b}$ is the bridge output voltage,  $\alpha_{\rm R} = 0.6\,{\rm ppm\,K^{-1}}$,  $\alpha_{\rm GIA} = 25\,{\rm ppm\,K^{-1}}$, $\alpha_{\rm C} = 30\,{\rm ppm\,K^{-1}}$, and  $\alpha_{\rm ADC} = 7\,{\rm ppm\,K^{-1}}$.}
\begin{center}
\begin{tabular}{l c c}
\hline\hline
\multirow{2}{*}{Source} & \multicolumn{1} {c}{${TC}_{\rm rti} = k_{\rm TC}\Delta R_{\rm b,n}$}   & \multirow{2}{*}{$k_{\rm TC}$} \\
            & \multicolumn{1} {c}{$[{\rm V/K}]$}  &  \\\hline
WB resistor & \multirow{2}{*}{$\alpha_{\rm R, AMR}V_{\rm b}/R_{\rm b}\Delta R_{\rm b}$}    & \multirow{2}{*}{$1.3 \times 10^{-5} $} \\ 
(volt. source)&   & \\
WB sensitivity  & $\alpha_{\rm sens}I_{\rm b}\Delta R_{\rm b}$    & $3.0 \times 10^{-6}$  \\
Bias source  & $\alpha_{\rm R}I_{\rm b}\Delta R_{\rm b}$    & $3.0 \times 10^{-9} $  \\
IA gain drift & $\alpha_{\rm GIA}I_{\rm b}\Delta R_{\rm b}$    & $1.3 \times 10^{-7} $  \\
IA offset   & \multirow{2}{*}{$(\alpha_{ \rm R}^2 + \alpha_{\rm C}^2)^{1/2}I_{\rm b}\Delta R_{\rm b}$}    & \multirow{2}{*}{$1.5 \times 10^{-7} $} \\ 
compensation  &  &    \\ 
Demodulator  & $\alpha_{\rm R}I_{\rm b}\Delta R_{\rm b}$    & $3.0 \times 10^{-9} $  \\
Integrator  & $(\alpha_{\rm R}^2 + \alpha_{\rm C}^2)^{1/2}I_{\rm b}\Delta R_{\rm b}$    & $1.5 \times 10^{-7} $  \\
Compensation  & \multirow{2}{*}{$\alpha_{\rm R}I_{\rm b}\Delta R_{\rm b}$}    & \multirow{2}{*}{$3.0 \times 10^{-9} $} \\ 
source  &  &    \\ 
ADC    & $\alpha_{\rm ADC}I_{\rm b}\Delta R_{\rm b}$    & $3.5 \times 10^{-8} $  \\\hline
${\rm Total_{V-Source}}$    & \multirow{2}{*}{$\sqrt{\Sigma TC_{\rm rti}^2}$}    & $1.3 \times 10^{-5} $  \\
${\rm Total_{I-Source}}$&  &  $3.0 \times 10^{-6} $	 \\\hline\hline

\end{tabular}
\end{center}

\label{tab:TCs}
\end{table}
\end{center}

\section{Results: low-frequency magnetic noise spectral density}\label{sec:results}

\subsection{Low-frequency noise: stray field measurements}

\begin{figure}[ht!]
\centering
   \includegraphics[width=1.0\columnwidth]{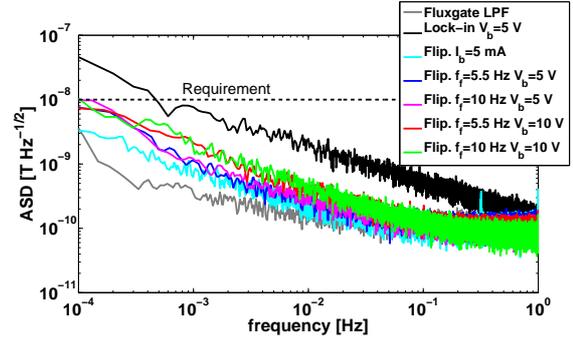}
   \caption {\label{fig:spectral_density} Equivalent magnetic field noise density for the engineering model of the fluxgate magnetometer used in LISA Pathfinder and AMR sensor using lock-in amplification and flipping techniques. Measurements have been done driving the AMR sensor with voltage ($V_{\rm b}$) and current source ($I_{\rm b}$, cyan trace). Bias field is not applied.}
\end{figure}

Low-frequency noise measurements for characterizing the system were carried out by placing the device inside a three-layer $\mu$-metal shielding. A bias field inside the shielding was not applied for these runs. Therefore, a low residual field around $ 20\,{\rm nT}$ was measured by the sensor. Lock-in and flipping noise reduction techniques at different modulation frequencies (5.5\,Hz and 10\,Hz) were performed by using  voltage ($5\,{\rm V}$ and $10\,{\rm V}$) and current sources ($5\,{\rm mA}$) to supply the sensor. Fig.\,\ref{fig:spectral_density} shows the equivalent magnetic field spectral density measured by the system. Noise curves exhibit that AC bridge excitation using lock-in amplification does not improve the potential $1/f$ noise of the magnetoresistors. Nevertheless, the flipping scheme helps to reduce part of it across the desired bandwidth. This contribution can not be totally mitigated and the noise measurements still exhibit a dominant $1/f$ behavior coming from the bridge's resistors. Thus, {\em excess} noise caused by the current that flows across the magnetoresistance bridge could be a significant contribution, which also exhibits a $1/f$ noise power spectrum\;\cite{Motchen,bib:1/f}. Owing to the TC reduction of the sensor, additional improvement in the millihertz bandwidth has been obtained when a low noise current source supplies the bridge instead of a voltage source,  showing a similar low-frequency noise to the fluxgate sensor used in LISA Pathfinder. At higher frequencies, the noise level is slightly reduced when flipping frequency and bridge voltage are raised. However, the increase in bridge voltage also implies higher {\em excess current} noise in the low-frequency band. Besides, more periodic flipping pulses can also induce additional disturbances in the spacecraft's environment. Since our interest is focused on the fluctuations in the low-frequency region, which is the limiting noise factor, flipping pulses at $5.5\,{\rm Hz}$ and a DC bridge current of $5\,{\rm mA}\, (V_{\rm b}= 4.25\,{\rm V})$ are the selected features.

\subsection{Low-frequency noise under a bias magnetic field}

Additional noise in the low frequency band appears as a consequence of the thermal dependence of the sensor coupled to the slow environmental temperature drifts. As shown in Table\,\ref{tab:TCs} the TC of the system depends on the amount of unbalance of the Wheatstone bridge, i.e., the magnitude of the magnetic field seen by the sensor. For noise investigation, a leading field of $\simeq 21\,{\rm \mu T}$ with a stability better than $1 \,{\rm nT \ Hz}^{-1/2}$ at $0.1\,{\rm mHz}$ is applied by a coil inside the magnetic shielding. The purpose is to unbalance the bridge to stress the effect of the gain temperature coefficient of the sensor and conditioning circuit during the noise measurements.

\begin{figure}[ht!]
\centering
   \includegraphics[width=1.0\columnwidth]{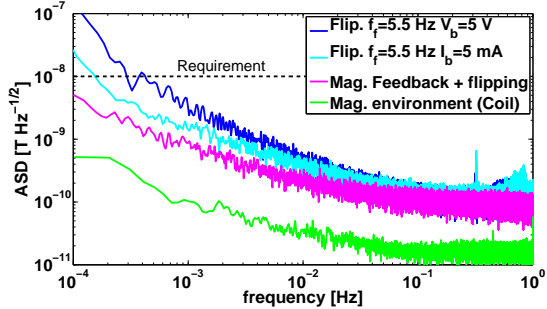}
   \caption {\label{fig:spectral_density_feedback_wmag} Spectral density in terms of equivalent magnetic field using flipping with voltage and current sources and electro-magnetic feedback. The green trace shows magnetic field noise generated by the coil at 15 mm from the sensor. The magnetic noise requirement at $0.1 {\rm mHz}$ (dashed trace) is achieved by using electro-magnetic feedback (magenta trace). Bias field is $\simeq 21\,{\rm \mu T}$.}
\end{figure}

Fig.\,\ref{fig:spectral_density_feedback_wmag} shows the noise measurements when using the flipping technique and electro-magnetic feedback in the presence of a bias field. To begin with, we measured the stability of the current source that generates the bias field and found it suitable to carry out the experiment. On the one hand, as far as the flipping method is concerned, the equivalent magnetic field noise at $0.1\,{\rm mHz}$ increases by an order of magnitude with respect to the previous results in Fig.\,\ref{fig:spectral_density} without a leading field. As expected, the additional noise due to thermal dependence is still more significant when a constant voltage source supplies the bridge.  On the other hand, the noise curve for electro-magnetic feedback shows that the effect of the gain temperature coefficient of the sensor is mitigated by the use of a proper closed-loop mode. The noise level achieved is $\simeq 5 \,{\rm nT \ Hz}^{-1/2}$ at $0.1\,{\rm mHz}$, which is below the requirement. Therefore, this compensation method is crucial to maintain long-term stability over temperature, and produces desirable results for measuring magnetic fields at the LISA frequencies. At higher frequencies, the noise floor is down to $\simeq 100\,{\rm pT\,Hz}^{-1/2}$ with the corner frequency around $ 0.2\,{\rm Hz}$. We remark that the noise floor might be reduced to the Johnson noise of the bridge resistance by increasing the flipping frequency.   

\section{Conclusions}\label{sec:conclusions}
We have presented the low-frequency noise characterization of a magnetic field measuring system based on AMR. Chip-scale magnetoresistive sensors appear as a solution to the disadvantages met when using the bulky fluxgate magnetometers in LISA Pathfinder. Nevertheless, magnetoresistors exhibit higher intrinsic noise characteristics than fluxgate magnetometers. Thus, in order to enhance the noise performance various methods have been analyzed and tested in the millihertz band. First, flipping techniques help to overcome part of the potential $1/f$ noise, which cannot be avoided with conventional lock-in amplification techniques(AC excitation of the bridge). Secondly, an excess noise below $1\,{\rm mHz}$ is exhibited when a magnetic field is applied to the sensor as a result of the temperature dependence of the sensor. A solution is found when using electro-magnetic feedback in the signal conditioning circuit. A closed-loop controller with a compensation coil helps to overcome the thermal dependence and to minimize the additional noise in the bandwidth of interest. With the combination of these methods, the equivalent magnetic noise spectral density is comfortably compliant with the envisaged noise requirement. Therefore, from the achieved noise performance, AMR sensors with dedicated noise reduction techniques are presented as an alternative to the fluxgate sensors used in LISA Pathfinder. With respect to the results published so far, we present a significant improvement of noise performance in the frequency range of the millihertz. Finally, we remark that the technology is likely be useful beyond the scope of eLISA, especially for space applications like STE-QUEST, with strict restrictions in size, weight, power, and low magnetic noise at low frequencies.



\section*{Acknowledgments}
We wish to thank the Paleomagnetic group in the Institute of Earth Sciences Jaume Almera (CSIC-UB) for the availability of their facilities and Antonia Morales-Garoffolo for her helpful input. Support for this work came from Project AYA2010-15709 of the Spanish Ministry of Science and Innovation (MICINN), ESP2013-47637-P of the Spanish Ministry of Economy and Competitiveness (MINECO), and 2009-SGR-935 (AGAUR).



\end{document}